\documentclass[]{aa}  
\usepackage{graphicx}
\usepackage{txfonts}
\usepackage{natbib}
\bibpunct{(}{)}{;}{a}{}{,} 

\def\lsxf{$L_{\rm{soft X}}/L_{\rm{FIR}}$}
\def\llsxf{$log(L_{\rm{soft X}}/L_{\rm{FIR}})$}
\def\lsx{$L_{\rm{soft X}}$}
\def\lx{$L_{\rm{X}}$}
\def\lfir{$L_{\rm{FIR}}$}
\def\xeff{$\epsilon_{\rm{xeff}}$}
\def\sfr{{\em SFR}}
\def\sfs{{\em SFS}}
\def\ergs{erg s$^{-1}$}
\def\msun{M$_\odot$}
\def\ebv{{\em E(B-V)}}

\begin{document}
\title{Soft X-ray to Far Infrared luminosities ratio in star-forming
galaxies: Predictions from synthesis models }

\titlerunning{Soft X-ray to Far Infrared luminosities ratio in star-forming
galaxies}

   \author{J.M. Mas-Hesse\inst{1,2}
          \and
          H. Ot\'{\i}-Floranes\inst{2,3}
	  \and
	  	  M. Cervi\~no\inst{4}
          }

   \offprints{J.M. Mas-Hesse}

   \institute{Centro de Astrobiolog\'{\i}a (CSIC--INTA), 28850 Torrej\'on de
Ardoz, Spain\\
              \email{mm@laeff.inta.es}
         \and
             Laboratorio de Astrof\'{\i}sica Espacial y F\'{\i}sica Fundamental
(LAEFF--INTA), POB 78, 
	     28691 Villanueva de la Ca\~nada, Spain
	     \and
	     Dpto. de F\'{\i}sica Moderna, Facultad de Ciencias, Universidad de Cantabria, 39005
	      Santander, Spain\\
             \email{otih@laeff.inta.es}
          \and
	     Instituto de Astrof\'{\i}sica de Andaluc\'{\i}a (CSIC), 18008
Granada, Spain\\
	     \email{mcs@iaa.es}
             }

   \authorrunning{Mas-Hesse et al.}
   
   \date{Received august, 2007; revised december 2007; accepted january 15th, 2008}
 
  \abstract
 {A good correlation has been found in star-forming galaxies, between the soft
X-ray and the far infrared or radio luminosities. The soft X-ray emission in
star-forming regions is driven by the heating of the diffuse interstellar medium, and
by the mechanical energy released by stellar winds and supernova explosions,
both directly linked to the strength of the star formation episode. }
{ We analyze the relation between the soft X-ray and far infrared
luminosities as predicted by evolutionary population synthesis models, 
aiming first  to test the validity of  the soft X-ray luminosity as a star
formation rate estimator, using  the already known calibration of the FIR
luminosity as a proxy, and second to propose a calibration based on the
predictions of evolutionary synthesis models.} 
{ We have computed the soft X-ray and far infrared luminosities expected for a
massive starburst as a function of { evolutionary state, the efficiency  of the
conversion of mechanical energy into soft X-ray luminosity, the star formation
history (instantaneous or extended) and dust abundance,}{} and we have compared these
predictions with observational values for 62 star-forming galaxies
taken from the literature. }
{The observational \lsxf\ ratios are consistent with the model predictions under realistic assumptions (young starbursts, and efficiency in the re-processing of mechanical energy of a few percent), confirming
the correlation between the diffuse soft X-ray emission and the star formation
episode.  }  
{The soft X-ray emission of the diffuse, extended gas surrounding massive star-forming regions, can be used as a star formation rate {tracer}.  The
empirical calibrations presented in the literature are supported by the
predictions of evolutionary synthesis models, and by the analysis of a larger
number of star-forming galaxies The calibrations are, however, biased towards galaxies dominated by
relatively unevolved starbursts. }


   \maketitle

\section{Introduction}

The onset of massive star formation episodes in galaxies drives their
observational properties in almost any wavelength range. The UV and optical
become dominated by the continuum of massive, hot and young stars, as well as by
the presence of nebular emission lines. After a few Myr of evolution, red
supergiant stars contribute to most of the near infrared emission. The heating
of interstellar dust particles by the powerful UV photons induces the thermal
re-emission of large amounts of energy in the { mid and far infrared} domain.
The injection of ionizing photons into the surrounding gas generates thermal
radio emission, which is replaced by non-thermal emission as the ionizing power
of the burst declines and the more massive stars begin to explode as supernovae.
 The direct relation between the strength of the star formation episode and the
intensity of the different observable parameters has allowed a number of star
formation rate calibrators to be defined, such as UV continuum, emission lines
intensity, far infrared or radio luminosities
\citep{Kennicutt98,Rosa02,Bell2003}. These calibrators have proven to be
invaluable for statistical studies of the star formation history of the
Universe.

Star-forming regions are also the source of  conspicuous X-ray emission,
generated by individual stars, by the injection of large amounts of
mechanical energy heating the interstellar medium, by supernova remnants, and by
binary systems transferring mass to a compact primary \citep{Cervino02,Persic02}. All of
these individual components are in principle  directly linked to the strength of
the star formation episode, so that the X-ray luminosity could also be used as
an estimator of star formation rates (\sfr). 

\begin{figure*}
\begin{center}
\includegraphics[width=8 cm,bb=5 39 786 521
dpi,clip=true]{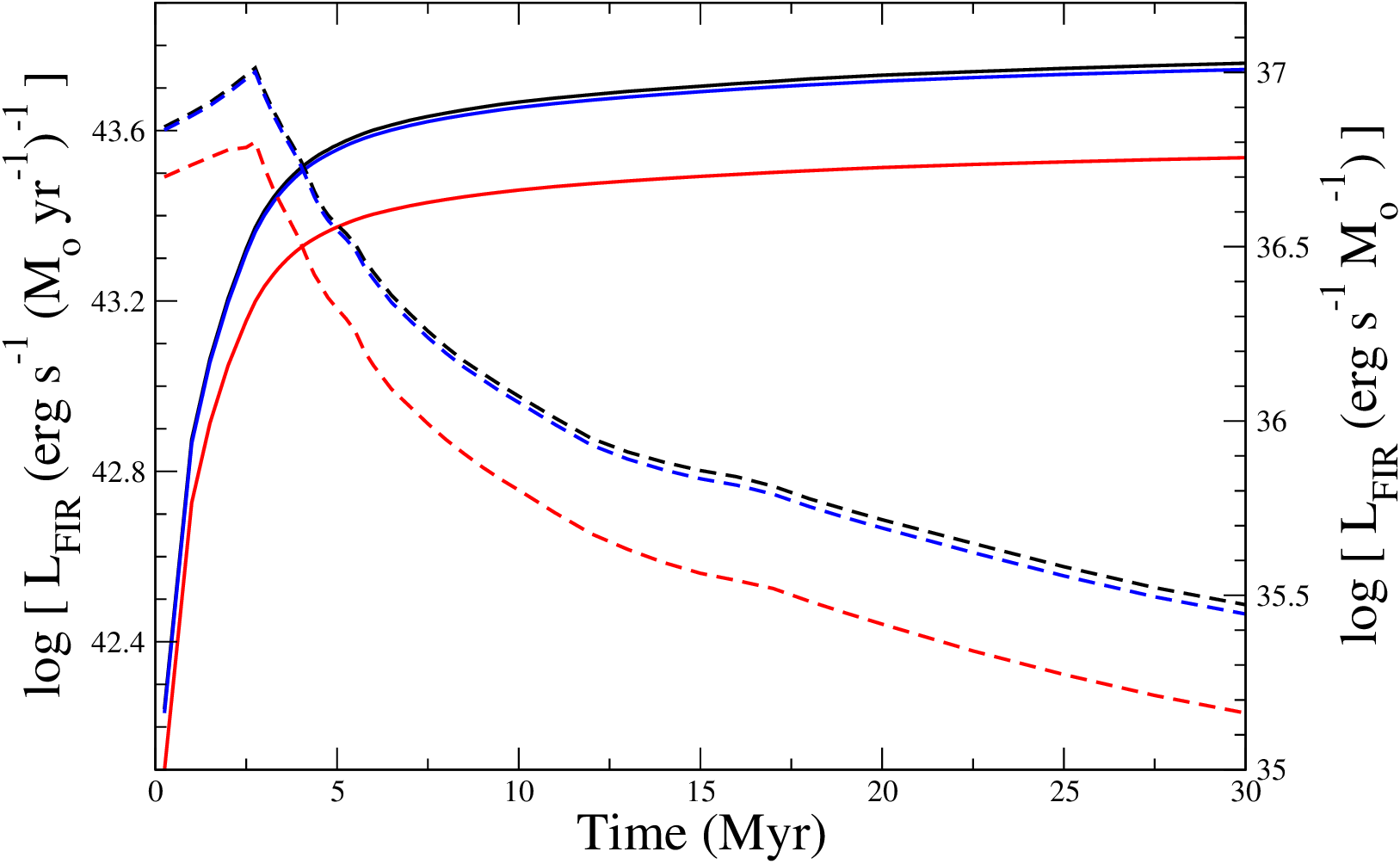}
\hspace{1.0 cm}
\includegraphics[width=8 cm,bb=17 39 786 527
dpi,clip=true]{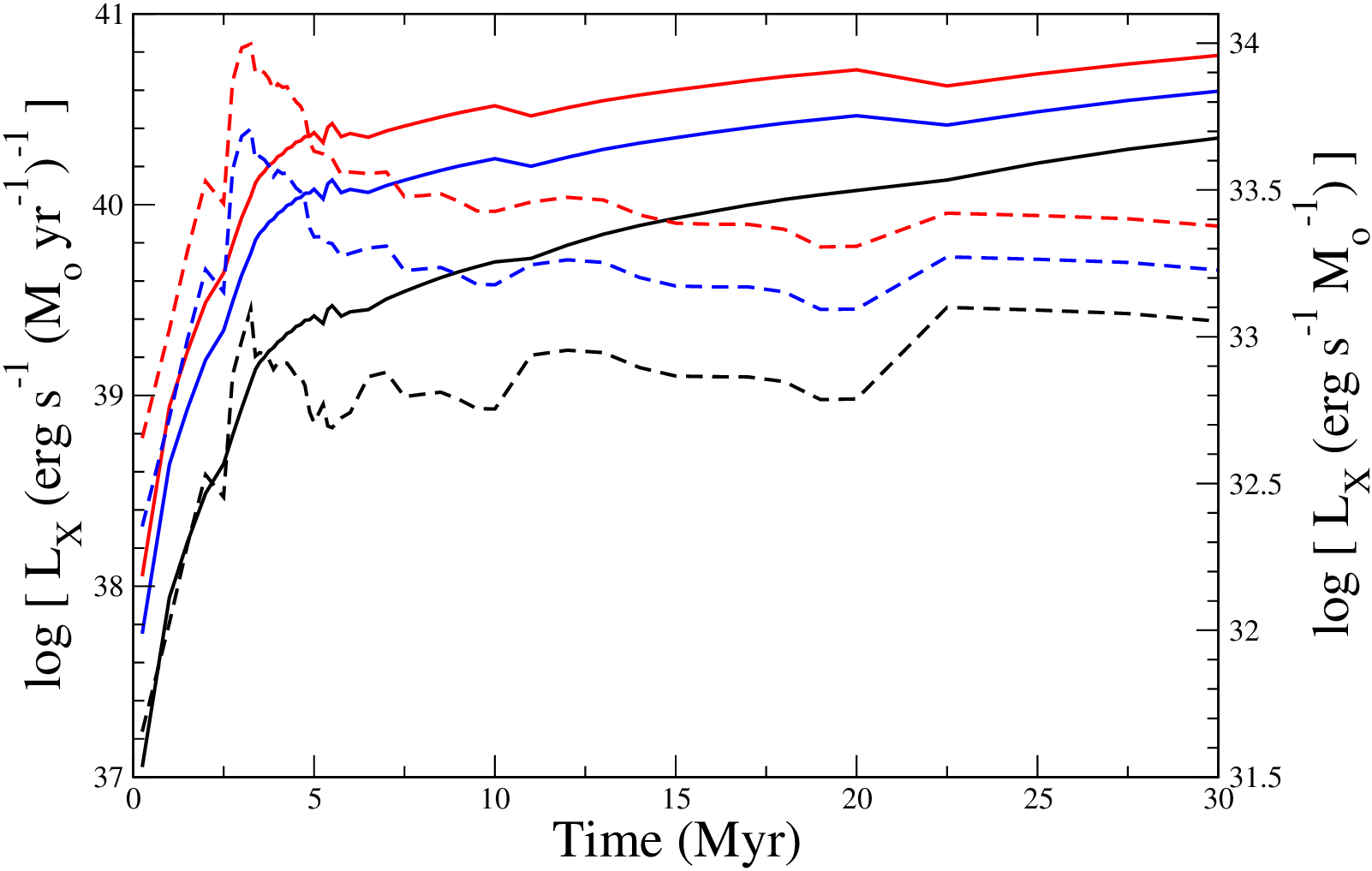}
\end{center}
\caption{Evolution of \lfir\ (left) and \lsx\ (right) for IB (dashed line, right
axis) and EB (solid line, left axis) models. IB models predictions are
shown normalized to $1$~\msun\ of gas transformed into stars, while for EB the
luminosities are scaled  to \sfr\ $= 1$ \msun\ yr$^{-1}$. \lfir\ has been
plotted (from top to bottom) for \ebv\ $= 1.0$, $0.5$ and $0.1$. \lsx\ has been
computed (from top to bottom) for \xeff\ $= 0.1$, $0.05$ and $0.01$.  }
\label{lumin}
\end{figure*}

Several authors have in recent years discussed the feasibility of using the
X-ray luminosity as an \sfr\ estimator. {\citet{Fabbiano02} already concluded
from the analysis of 234 S0/a-Irr galaxies observed with {\em Einstein}, that
the correlation they found between the X-ray and the FIR luminosities in Sc-Irr
galaxies was due to the young stellar populations in these objects.}
\citet{Ranalli03} proposed an empirical calibration of the soft ($0.5$--$2.0$
keV) and hard ($2$--$10$ keV) X-ray luminosities, based on their correlation
with the far infrared (FIR) and radio luminosities, and using the known 
calibrations of these parameters as proxies. \citet{Grimm03} studied the
correlation between the number of high-mass X-ray binaries (HMXB) and the \sfr,
deriving different calibrations of the hard X-ray luminosity for low and high
star formation rates.  \citet{Persic04} obtained a calibration of the hard X-ray
luminosity as a \sfr\ estimator by assuming that most of the emission in this
range is associated with HMXB, and scaling from the number of HMXB to the \sfr\
of our Galaxy. \citet{Gilfanov2004} confirmed the {calibration} of the hard
X-ray luminosity associated with HMXB, using slightly different slopes at high
and low star formation rates.  In a recent paper, \citet{Persic07} found indeed
that the collective hard X-ray emission of young point sources correlates
linearly with the star formation rate derived from the far infrared luminosity.
\citet{Strickland04b} demonstrated that the luminosity of diffuse X-ray emission
in star-forming galaxies is directly proportional to the rate of mechanical
energy injection from the young, massive stars into the interstellar medium of
the host galaxies. A similar result was found by \citet{Grimes05} from the
analysis of  a sample of starburst galaxies of different types  (from dwarf
starbursts to ultraluminous infrared galaxies), which concluded that the
mechanism producing the diffuse X-ray emission in the different types of
starbursts was powered by the mechanical energy injected by stellar winds and
supernovae into the surrounding medium.  Recently \citet{Rosa07} confirmed the
reliability of the soft X-ray luminosity as an \sfr\ estimator from the analysis
of a sample of star-forming galaxies in the {\em Chandra Deep Field South} at
redshifts $ z = 0.01 - 0.67$, using the UV continuum luminosity from {\em GALEX}
as a proxy.

While hard X-ray emission from star-forming regions may be dominated by binary
systems, diffuse soft X-ray emission is generated by reprocessed, mechanical
energy from stellar winds and supernovae explosions. This mechanical energy is
related to the strength of the burst of star-formation, and can be calculated
using evolutionary population synthesis models. In this paper, we analyze the
correlation between the soft X-ray and FIR luminosities in star-forming regions,
both predicted by evolutionary synthesis models. We study the \lsxf\ ratio as a
function of the {evolutionary state, the efficiency  of the conversion of
mechanical energy into soft X-ray luminosity, the star formation history
(instantaneous or extended), and the dust abundance,} and compare the computed
values with observations taken from the literature. { Our objective is to derive
a calibration of \lsx\ as a tracer of the star formation rate, based on the
predictions of evolutionary synthesis models, and to  test the validity of the
empirical calibration proposed by \citet{Ranalli03}.  } 

In Sect.~2 we describe the evolutionary synthesis models that we use in the present study, in Sect.~3 we
present the observational data taken from the literature, and in Sect.~4 we
discuss the predictions and the comparison with the observational values.  {\
Throughout this work we have assumed $H_{\rm0} = 73$ km s$^{-1}$ Mpc$ ^{-1}$ to
convert fluxes into luminosities.}

\begin{figure*}
\begin{center}
\includegraphics[width=8cm,bb=20 39 713 527
dpi,clip=true]{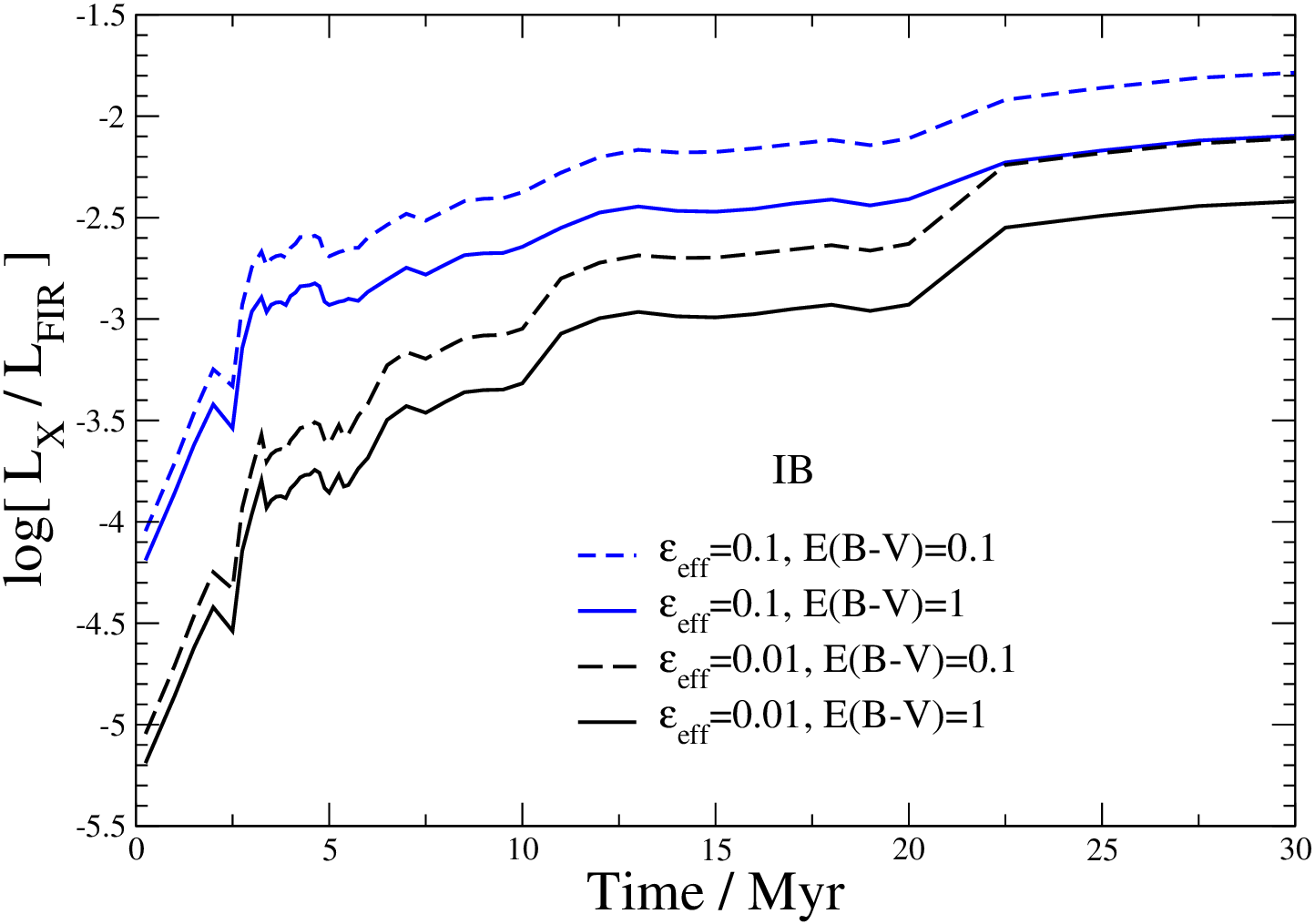}
\hspace{1.0 cm}
\includegraphics[width=8cm,bb=20 39 713 527
dpi,clip=true]{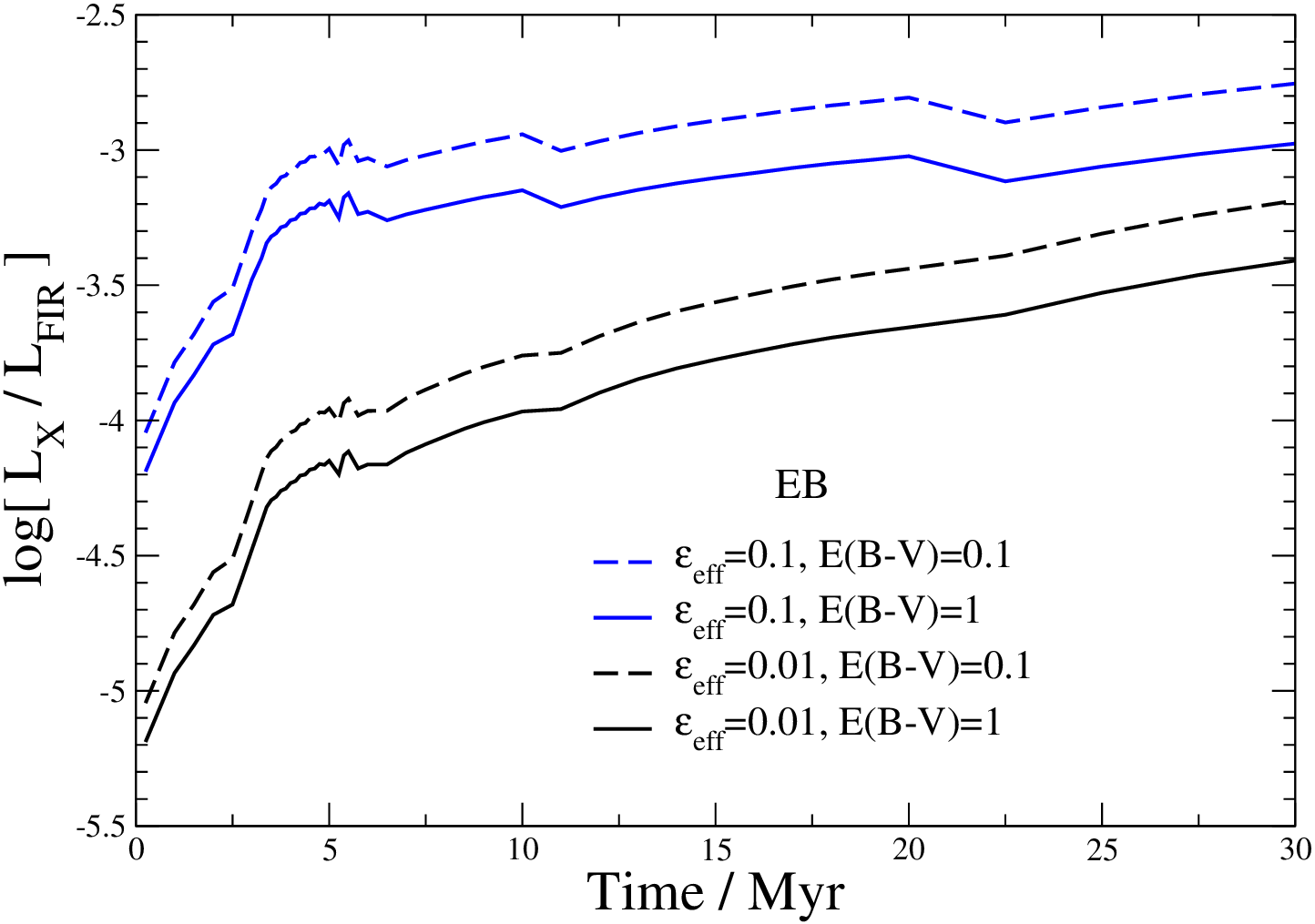}
\end{center}
\caption{Evolution of the \lsxf\ ratio computed for \xeff\ $= 0.01, 0.1$ and \ebv\ $= 0.1,  1.0$. Left: predictions for IB models; right: predictions for EB models.  } 
\label{ratio}
\end{figure*}
 
\section{Evolutionary synthesis models}

We have computed the predicted \lsx\ and \lfir\ values using the evolutionary
population synthesis models of \citet{Cervino02} (hereafter CMHK02 models\footnote{
Downloadable from {\rm http://www.laeff.inta.es/users/mcs/SED/}}), which are
based on the models of \citet{Arnault89}, \citet{MasHesse91} and
\citet{Cervino94}. These models compute the evolution of a cluster of massive
stars, formed at the same time (Instantaneous Bursts, IB), or {during an extended
period of time (typically several tens of Myr) at a constant rate (Extended
Bursts, EB)}, assuming different metallicities and Initial Mass Function (IMF)
slopes. { The CMHK02 models were developed to compute the evolution of a
starburst during the first 30~Myr, following the onset of the star formation
episode.} Once the structure {(number of stars of each spectral type and
luminosity class at a given evolutionary time)} of the stellar population is
derived, the models are used to compute a number of observable parameters, from X-ray to radio wavelengths.  

In general, we assume a solar metallicity $Z_{\sun}$, and a Salpeter IMF ($\phi(m)\sim
m^{-2.35}$) with masses ranging between $2 \, M_{\sun}$ and $120 \, M_{\sun}$. {The output of the models are normalized to the star formation rate (mass of gas
transformed into stars per unit time, $M_{\sun}  yr^{-1}$) for EB cases, or star
formation strength (mass of gas transformed into stars at the onset of the
burst, in units of $M_{\sun}$) for IB scenarios. In both cases, the mass
normalization corresponds to the mass integrated between 2 $M_{\sun}$ and 120 $M_{\sun}$
assuming a Salpeter IMF. We emphasize that this normalization might differ
significantly if other mass limits are considered. For example, the ratio
between our mass normalization ($2-120 M_{\sun}$) and the one assumed by
\citet{Kennicutt98}  ($0.1-100 M_{\sun}$) is $M_2^{120}/M_{0.1}^{100} = 0.293$.
It is important to remark that this is the normalization implicitly assumed when
the \sfr\ calibrations proposed by \citet{Kennicutt98} are used.}  Under our
assumptions, the soft X-ray ($0.4$ -- $2.4$~keV) and far infrared luminosities
were computed during the first 30~Myr after the onset of the massive starburst
episode.

One of the main sources of soft X-ray emission in a star-forming region is the
diffuse gas heated by mechanical energy from a starburst (massive
stellar winds or supernova explosions) into the surrounding medium. Its
contribution  is modeled by a Raymond-Smith thermal plasma with $kT=0.5$ keV,
which controls the fraction of the mechanical energy that heats the gas to X-ray temperatures, \xeff.  The models include the soft X-ray
radiation emitted during the adiabatic phase of the Supernova Remnant (SNR),
which is modeled by a composite Raymond-Smith plasma with $kT=0.23$, $0.76$ and
$1.29$ keV.  The total energy emitted by the SNR, during the adiabatic phase, has
been subtracted from the energy of each supernova explosion when computing the
injection of mechanical energy.  A more detailed description of our models is provided in \citet{Cervino02}.

The contribution of the stellar atmospheres to the soft X-ray emission was
neglected because it is expected to be two orders of magnitude lower than the
emission from the diffuse gas. The contribution of high-mass X-ray binaries to
the soft X-ray emission has in addition been neglected in this work.
\citet{MasHesse99a} discussed the properties of the HMXB population expected to
form during a massive star-formation burst. Binary systems become X-ray emitters
when the primary collapses into a black hole or neutron star, the atmosphere of
the secondary has started to expand, and the secondary is sufficiently close to
the collapsed primary for mass transfer to begin. Mass is accreted onto the
surface of the compact object and emits X-rays with a typical \lx\ $\sim
10^{38}$ \ergs, peaking at energies between $5$ and $10$ keV \citep{Persic02}.
{The number of HMXB in a young starburst is dependent on many free parameters.
Following \citet{MasHesse99a}, we estimate that only a few HMXB should be active
after the first 5--6 Myr of evolution of starbursts that have transformed
approximately $10^6$ \msun of gas into stars. This HMXB population should
contribute a few times  $10^{38}$ erg s$^{-1}$ to the total X-ray luminosity,
and a small fraction of this radiation to the soft X-ray emission.} In all
cases, the total contribution of HMXB, to the soft X-ray emission, remains
$15$\% for IB and $10$\% for EB. Only if one or a few of these HMXB develop into
an ultraluminous X-ray source (ULX), with \lx\ $\sim 10^{40}$ \ergs
\citep{Miniutti06}, could the X-ray emission, from soft to hard X-rays, of the
entire galaxy, be dominated by the emission of HMXB, compared to that of the
diffuse gas. 

Concerning the FIR emission, a thermal equilibrium of dust is assumed, implying
that all the energy absorbed by dust, mostly originating from the UV continuum
of the  massive stars, is reemitted in the FIR range. { In this paper, \lfir\
refers to the total mid- and far-infrared luminosity integrated over the
wavelength range $1 -1000 ~\mu$m. We remark that this parametrization of \lfir\
implies a value of \lfir\ that is larger than that calculated using the FIR
parameter proposed by \citet{Helou88}, which is widely used in the literature.
\citet{Helou88} showed that the FIR luminosity computed from the {\em IRAS}
fluxes at $60$ and $100$ ~$\mu$m would intercept about 70\% (0.15 dex) of the
total FIR luminosity from 1 to 1000 $\mu$m, assuming that there is a single,
dominating component, at a temperature of 30 to 50 K.  The discrepancy would be
even larger in the presence of an additional warm dust component. This estimate
of \lfir\ is consistent with the range  ($8 -1000 ~\mu$m) considered by
\citet{Kennicutt98} for the calibration of \lfir\ as an \sfr\ estimator, since
most of the FIR luminosity in starburst galaxies is emitted at wavelengths in
the range $10 -120 ~\mu$m. The models apply a Galactic extinction law
\citep{Cardelli89} to the synthetic spectral energy distributions, which is
parametrized by the colour excess \ebv. The energy absorbed due to extinction is
calculated using the models. It is assumed that all energy absorbed is reemitted
thermally by the dust, within the mid to far infrared domain, i.e., within the
$\sim$ $8-1000 ~\mu$m range. The models do not predict the shape of the FIR
emission, since we do not make any assumption about the expected dust
temperature. The presence of completely-obscured stars is not taken into
account. Furthermore, the models assume that a fraction $1-f$ of Lyman continuum
photons is directly absorbed by the dust, and does not contribute to the
ionization (see \citet{Mezger1974}). {\citet{Mezger1978} for the Galactic and
\citet{Degioia1992} for the Large Magellanic Cloud HII regions, derived $1-f$
values in the range $0.3-0.4$.} We have assumed $(1-f)=0.3$ in this work, as
proposed for starburst galaxies by \citet{Belfort1987}. }
 
The evolution of both \lfir\ and \lsx\ predicted by the models are shown in
Fig.~\ref{lumin}, while their ratio \lsxf\ is presented in Fig.~\ref{ratio}. For
IB models, the luminosities are shown scaled to $1$~\msun\ of gas transformed into
stars at the onset of the starburst.  In the case of EB models, the luminosities
are normalized to  \sfr\ = $1 \, M_{\sun} \, yr^{-1}$. \lfir\ is presented in the
plot, computed for \ebv\ $= 0.1$, $0.5$ and $1.0$. It can be seen that FIR
emission saturates rapidly for \ebv\ values {above $0.5$.}  In the remainder of this work, we consider the value of \lfir\ calculated by assuming that \ebv\ $=
1.0$, which can be considered an upper limit to the expected FIR luminosity.
In the case of EB models, \lfir\ reaches an
asymptotic value after approximately $10 - 15$~Myr of evolution, when an equilibrium is
reached between the number of massive stars that die, and those forming
continuously. For coeval starbursts, \lfir\ declines rapidly after the
first $5$~Myr of evolution, when the most massive stars begin to end their
lifetimes and stop heating the interstellar dust. 

\begin{figure}
\centering
\includegraphics[width=8cm,bb=20 39 713 527 dpi,clip=true]
  {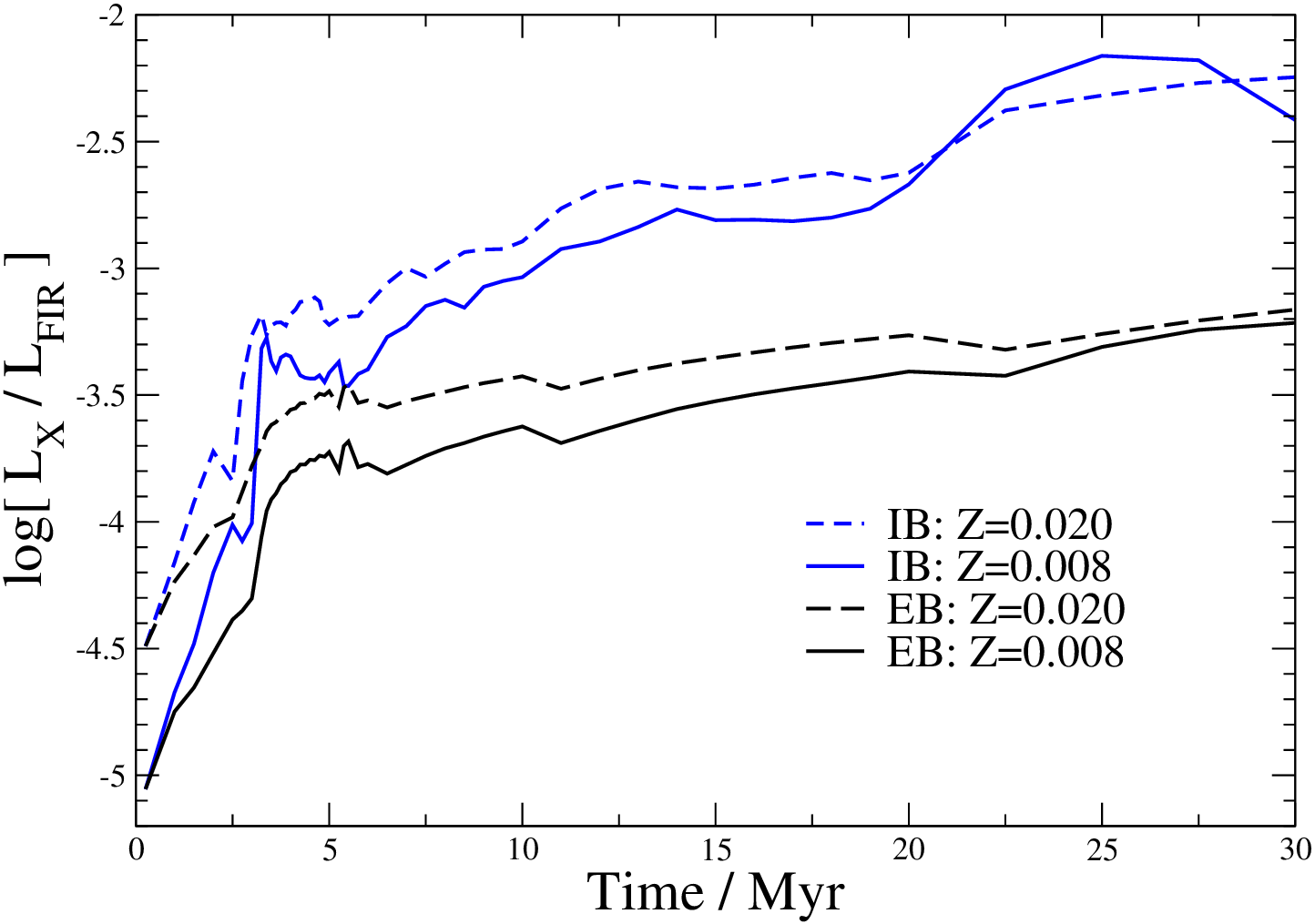}
\caption{{ Evolution of the \lsxf\ ratio as a function of metallicity and star
formation regime. \ebv$ =1.0$ and \xeff\ $=0.05$.}} 
\label{ratio-met}
\end{figure}

\lsx\ is shown in Fig.~\ref{lumin} computed for \xeff\ values of $1$\%, $5$\%
and $10$\%. During the first few Myr, there is a rapid increase in \lsx\ because
both the luminosity and the stellar winds of the most massive stars, increase. After the first $3$~Myr of evolution, the most massive stars end
their lifetimes and explode as supernovae. The injection of mechanical energy
begins to be dominated by energy released by supernovae explosions, as the
importance of stellar winds rapidly diminishes. { During the first 35 Myr, this
remains the situation for IB models because, for a Salpeter IMF, the supernovae
rate declines slowly, while there exist stars of sufficient mass (of initial mass
above $8$~\msun) to produce supernovae \citep{Cervino94,Leitherer95}. In the
case of EB models, \lsx\ is expected to increase slowly after the first 5~Myr,
until an equilibrium is reached between the formation and destruction of stars
that end their lifes as supernovae, i.e., at around $40$~Myr at solar
metallicities, according to the evolutionary tracks considered. 
\citet{Leitherer95} presented the evolution of both supernova rate and the
injection rate of mechanical energy for longer term evolution, up to ages of
300~Myr. Their Fig.~56 showed that the asymptotic rate of energy injection is
within 0.05 dex of the value predicted at 30~Myr}. Although the computation of
the soft X-ray luminosities in the CMHK02 models is simplistic, the
predictions are in good agreement with the results of \citet{Strickland99},
which were computed using hydrodynamical simulations of a young superbubble
driven by a cluster of massive stars.        

\begin{figure}
\centering
\includegraphics[width=7cm,bb=50 34 705 521 dpi,clip=true]
  {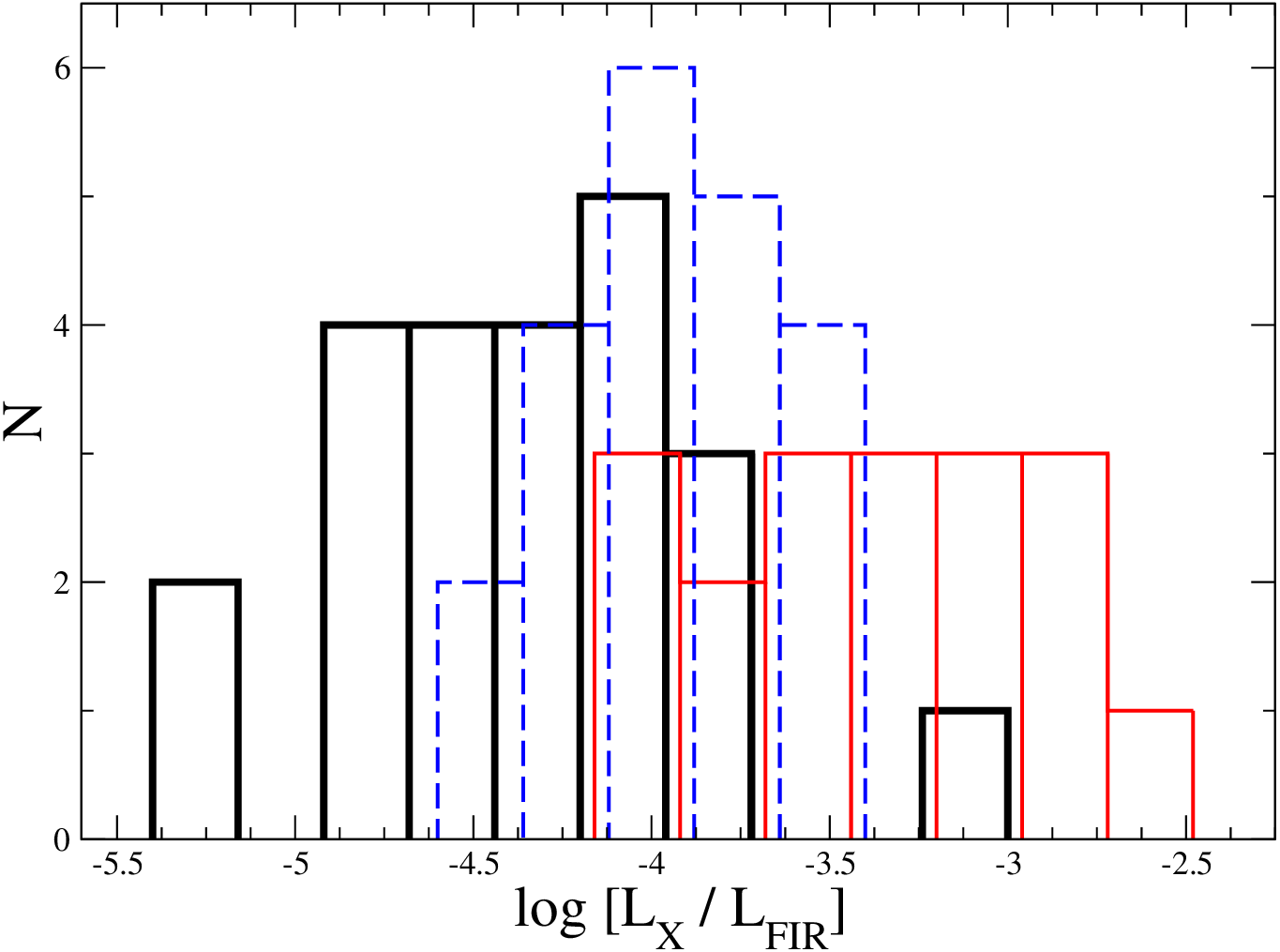}
\caption{\lsxf\ histograms for the samples of star-forming galaxies from
\citet{Ranalli03} (dashed line), \citet{Tullmann06b} (thick solid line) and
\citet{Rosa07} (thin solid line). { The bins have been computed with a slight
shift between each other for clarity.} } 
\label{histo} 
\end{figure}

Figure~\ref{ratio} shows the predictions for the \lsxf\ ratio evolution
with time.  It can be seen that \lsxf\ increases continuously with time after an
instantaneous burst, even after the first $5$~Myr, due to the fact that \lsx\
remains almost constant while \lfir\ decreases rapidly.  In extended burst
models, {the rate of increase of \lsxf\ with time,} after the first $5$~Myr, is smaller. As
discussed above,  we expect the \lsxf\ ratio to stabilize after about $40$ Myr
in EB models, when an equilibrium between formation and destruction of stars
susceptible to becoming supernovae, has been reached.  In both cases, there is a
rapid increase in the \lsxf\ ratio after the onset of the burst (one order of
magnitude in $3$~Myr in IB models), associated with the rapid increase in the
amount of mechanical energy injected during the first phase of the evolution of
the most massive stars.  

{ In Fig.~\ref{ratio-met}, we plot the evolution of the \lsxf\
ratio as a function of metallicity ($Z = Z_\odot$ and  $Z = 0.4\times Z_\odot$)
and star formation history (instantaneous and extended bursts). Varying metallicity highlights two results: first, low-metallicity stars evolve more slowly and
have a longer lifetime, such that the evolution of \lfir\ is delayed with respect
to solar-metallicity stars. Second, the lower the metallicity, the less
efficient are the stellar winds, and therefore the lower is the amount of mechanical
energy released into the interstellar medium. As seen in Fig.~\ref{ratio-met}, the net
effect is a decrease in the value of \lsxf\, at intermediate ages by 0.2 dex, or by up to 0.5 dex within the first 4 Myr of evolution. At some ages, the trend is even reversed. For the time interval considered, however, the predicted values of \lsxf\ are similar for both values of metallicity.  }

\begin{figure}[ht]
\centering
\includegraphics[width=8cm, bb= 19 39 705 565
dpi,clip=true]{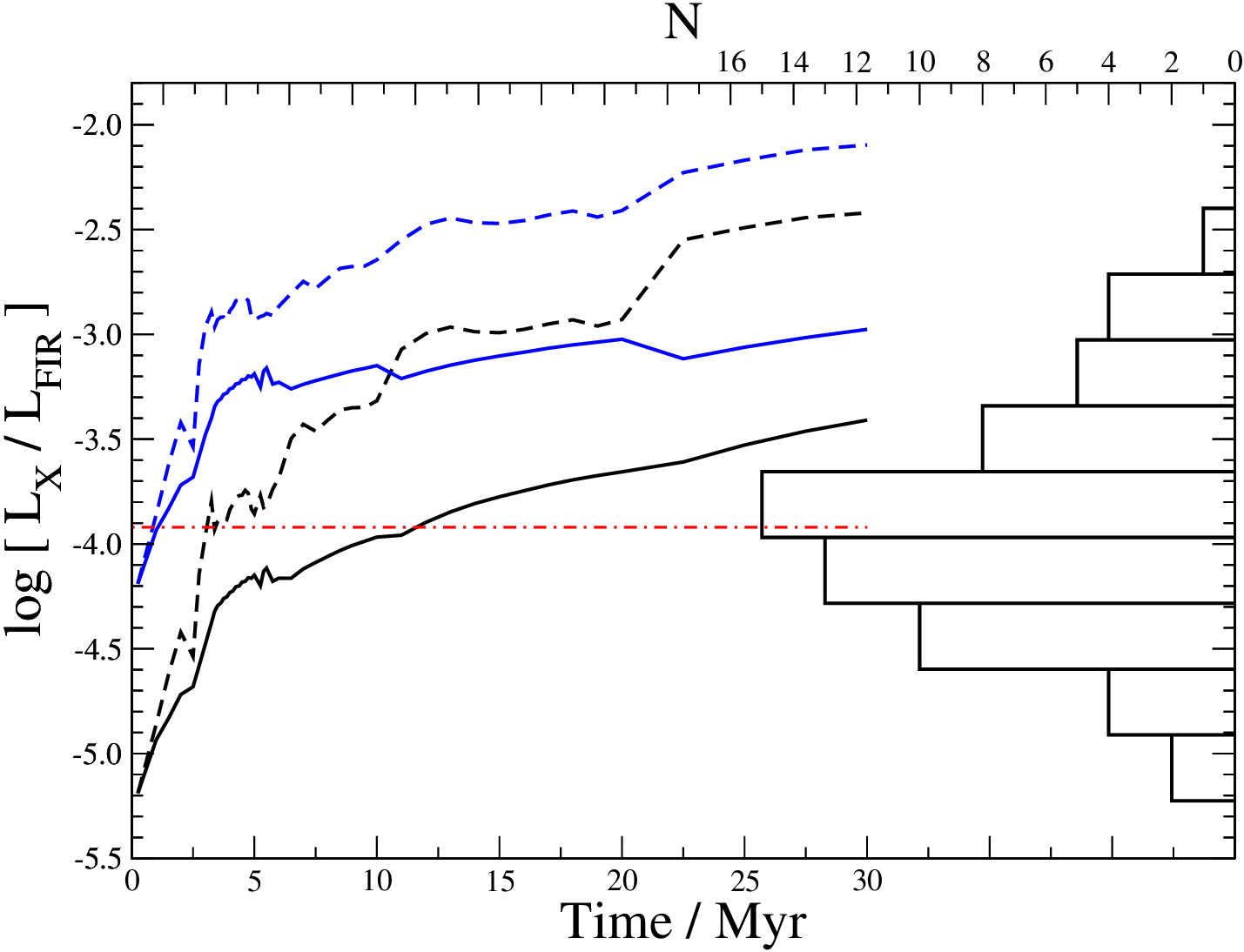}
\caption{\lsxf\ histogram for the combined sample, over the evolution of
the \lsxf\ computed for IB (dashed lines) and EB (solid lines) and \xeff\ $=
0.10$ (top) and $0.01$ (bottom). The horizontal line corresponds to the mean
ratio of the complete sample, \llsxf\ $= -3.92$.}
\label{sxf-histo}
\end{figure}
   
\section{Observational data sample}

We have compiled \lsx\ and \lfir\ data for 62 star-forming galaxies, of
different types and redshifts, to compare with our model predictions.

{The data compilation of \citet{Ranalli03} (hereafter RCS03) is for star-forming
galaxies, from the atlas of \citet{Ho97}, that have detectable X-ray emission in
ASCA and/or BeppoSAX observations. Only spiral and irregular galaxies from Sa to
later types were included in this sample, which was complemented by the authors
with data of six, well-known starburst galaxies observed in the southern
hemisphere.}

The sample compiled by \citet{Tullmann06b} (hereafter TUL06) is based on the nine
late-type starburst galaxies of \citet{Tullmann06a} ({\em XMM-Newton} data), 
seven star-forming disk galaxies from \citet{Strickland04a} ({\em Chandra} data)
and seven additional late-type star-forming galaxies taken from the literature.
The contribution by obvious point-sources within the extraction regions was
removed from all objects in the original \citet{Tullmann06a} and
\citet{Strickland04a} compilations when possible. 

{ In both samples, the X-ray emission was corrected for Galactic neutral
Hydrogen absorption, but not for the intrinsic absorption of the galaxies. The
published far-infrared fluxes  were computed using the {\em IRAS} 60 and 100
$\mu$m flux following the procedure of \citet{Helou88}, corresponding to the energy emitted
within the range $40 - 120 ~\mu$m. \citet{Calzetti2000} found for local starburst galaxies that $F(1-1000)/F(40-120) = 1.75 \pm 0.25$. We have
therefore multiplied the FIR fluxes provided by RCS03 and
TUL06 by 1.75, in order to obtain a more realistic determination of
the total amount of energy being reemitted in the mid- and far- infrared range.
The luminosities were recomputed for all objects from the published  fluxes
by using the distances corrected to the Local Group reference frame as given in
the {\em Nasa Extragalactic Database} , assuming   $H_{\rm0} = 73$ km s$^{-1}$
Mpc$ ^{-1}$. 

There is some overlap between the two samples. The \lsxf\ ratio for M82 is
$-4.08$ in RCS03 (based on {\em BeppoSAX} data) and $-3.85$ for
TUL06 ({\em Chandra}).  For NGC~4631 \lsxf\ varies between $-3.96$
({\em ASCA}) and $-4.17$ ({\em XMM-Newton}), respectively. In these cases we
have taken the \lsx\ values provided by  TUL06 . }

To compare with star-forming galaxies outside the Local Universe, we consider
galaxies at redshifts $ z = 0.01 - 0.67$ studied by \citet{Rosa07} (hereafter
ROSA07). Both \lsx\ and \lfir\ data are available for these galaxies, which were
observed as part of the Chandra Deep Field South (CDFS) survey. The X-ray data
in this sample come from the catalog of \citet{Alexander03}, and were not
corrected for either Galactic or intrinsic absorption by neutral hydrogen.
Nevertheless, along the line of sight to the {\em CDFS} the expected Galactic
neutral hydrogen absorption in the soft X-rays band is only 4.2\% (0.02
dex) \citep{Alexander03}. { We exclude objects believed to harbour an obscured
AGN, or to be dominated by low-mass, X-ray binaries (LMXB). In total, we have
identified 18 objects with detectable soft X-ray, and far-infrared fluxes that
appear not to be contaminated by either an AGN or LMXB. }  

ROSA07 derived the far infrared luminosities for these objects in the
full 8 -- 1000 $\mu$m band using Spitzer observations at 25 $\mu$m. They used the
empirical calibration of \citet{Takeuchi05} {based on the analysis of a
large sample of galaxies for which fluxes in the four {\em IRAS} bands at 12,
25, 60 and 100 $\mu$m were available.  The FIR luminosity computed in this way
should be consistent with the luminosities derived for the RCS03
 and TUL06 samples, extrapolated to the 1 -- 1000 $\mu$m range. 
ROSA07 assumed   $H_{\rm0} = 70$ km s$^{-1}$ Mpc$ ^{-1}$ to convert
fluxes into luminosities. } 

We note that the \lsx\ values are integrated over the $0.5 - 2.0$ keV band by
RCS03,   $0.2 - 2.0$ by ROSA07, and over the $0.3 -
2.0$ keV band by TUL06, while the model predictions were computed
for the $0.4 - 2.4$ keV band. We have verified that for the typical spectral
properties of the diffuse gas in these objects, the discrepancies associated with
the different bandwidths should be smaller than $9$\% ($0.04$ dex) in any case.

We have plotted in Fig.~\ref{histo} the histograms of the \lsxf\ distribution
for each sample. It can be seen that while there is a significant overlap
between them, the star-forming galaxies compiled by RCS03 show the
smallest dispersion. The TUL06 galaxies show generally lower \lsxf\
values than the objects compiled by RCS03, while the sample of
ROSA07 presents the highest \lsxf\ ratios. The mean \llsxf\
values derived for each sample are $-3.94$ (0.27) (RCS03), $-4.34$
(0.48) (TUL06), $-3.37$ (0.45) (ROSA07) and $-3.92$
(0.57)  for the whole compilation. Values within parentheses correspond to the
$\sigma$ dispersion of each sample. { We have looked for any possible
correlation between the \lsxf\ ratios in the TUL06 galaxies and
their morphological type,  but there is no clear trend. Both Sc+Sd and SB galaxies in the sample cover a wide range of luminosities and \lsxf\
values and furthermore, they cover a similar range in luminosities than the
galaxies in the RCS03 sample. There is no obvious reason for the
differences between the two samples of local star-forming galaxies.  }    

\section{Discussion}

Figure~\ref{ratio} indicates that our models predict a strong dependence of the \lsxf\
ratio on the star formation history (instantaneous or extended), and on the
evolutionary state of the star formation process. Moreover, the ratio is also
strongly dependent on the efficiency of the reprocessing of mechanical energy
and UV photons into soft X-ray and FIR emission, respectively. Therefore, we
would expect a significant scatter in the \lsxf\ values observed in star-forming
galaxies. This is indeed what we find in Fig.~\ref{histo}, as discussed above. 

\begin{figure*}
\begin{center}
\includegraphics[width=8 cm,bb=18 22 705 527
dpi,clip=true]{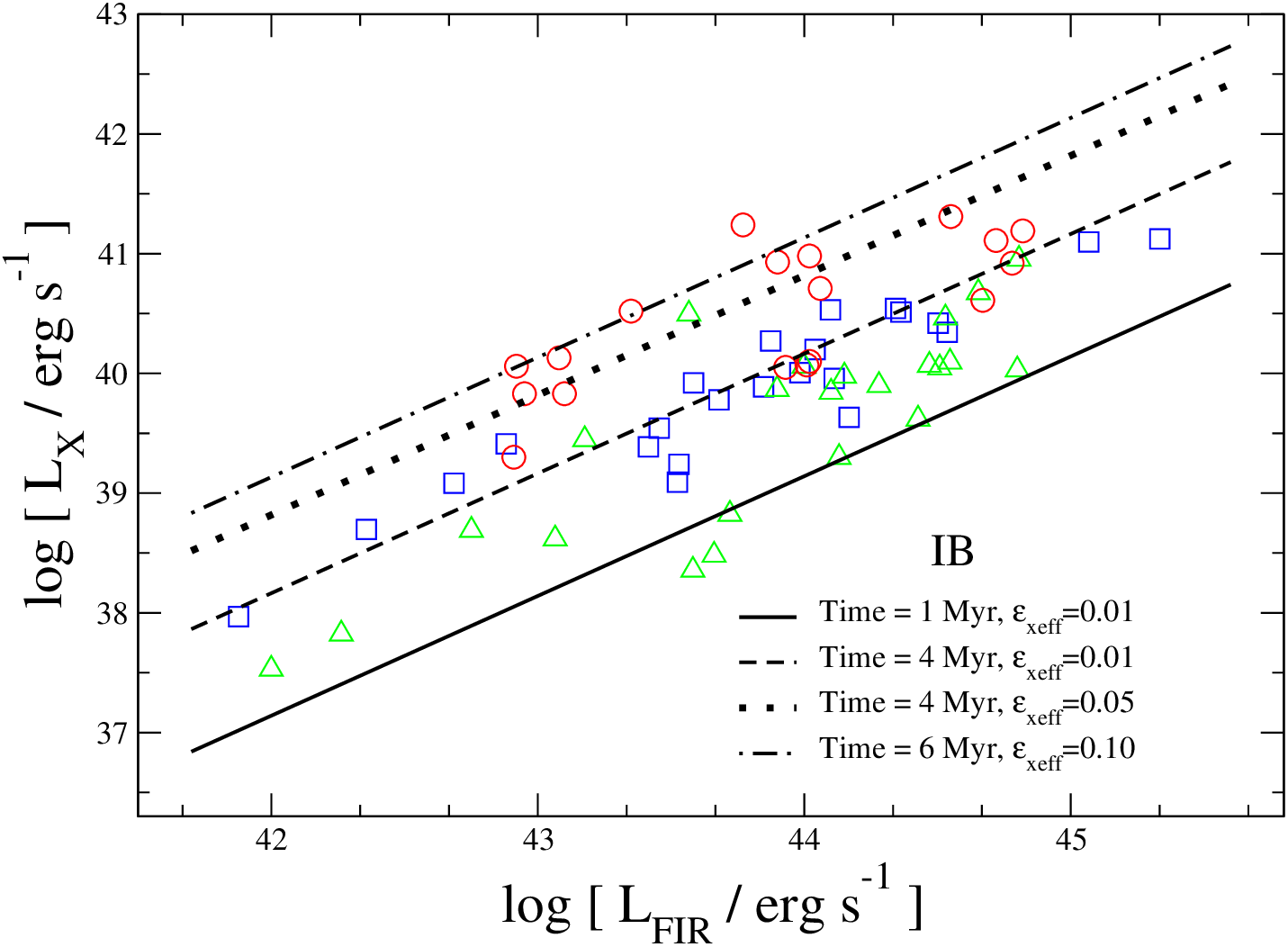}
\hspace{1.0 cm}
\includegraphics[width=8 cm,bb=18 22 705 527
dpi,clip=true]{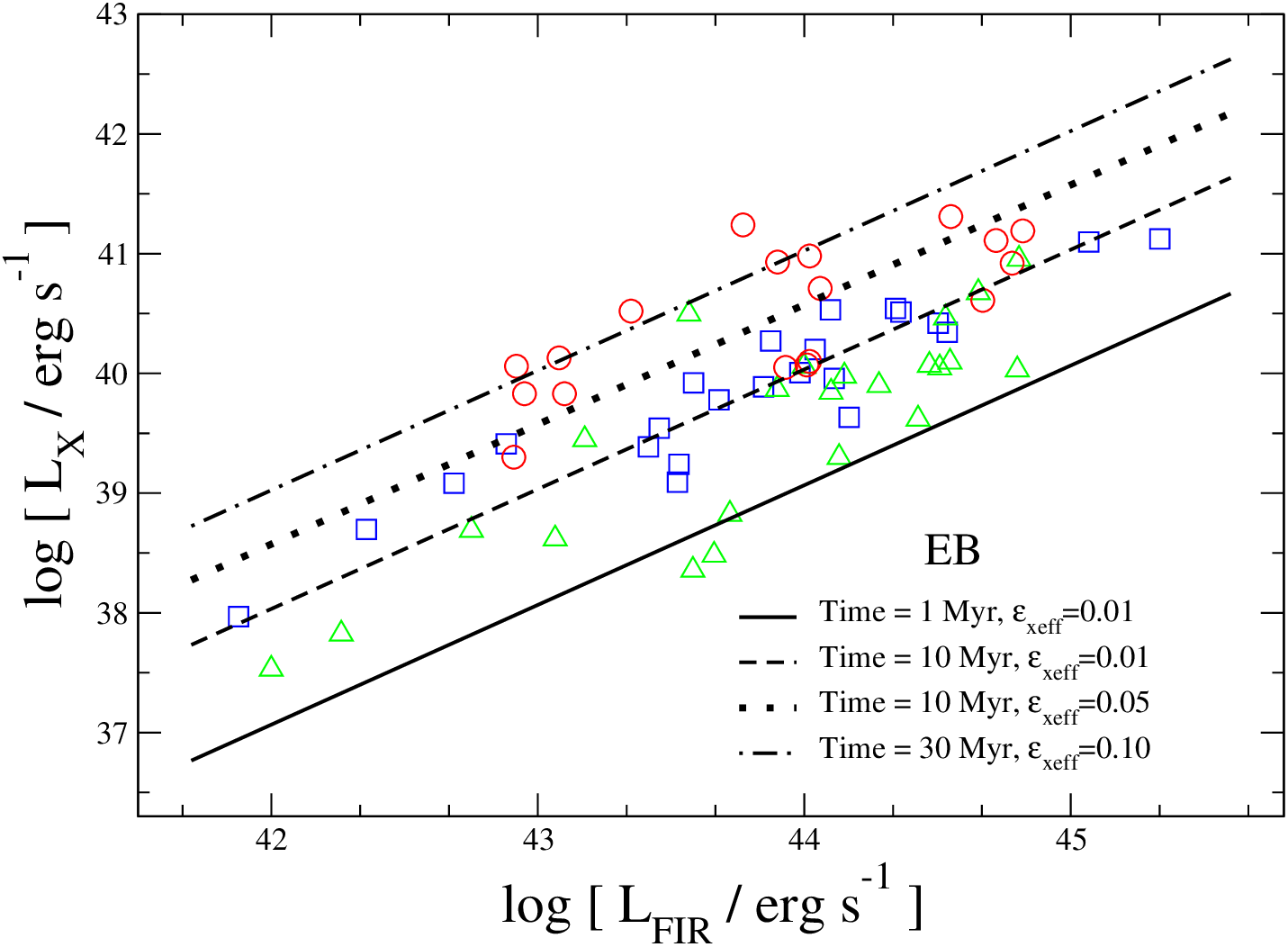}
\end{center}
\caption{\lsx\ vs. \lfir\ for the RCS03 (squares), 
TUL06 (triangles) and ROSA07 (circles) samples. 
In the left panel  we have
overplotted the correlation lines corresponding to the \lsxf\ ratios predicted
by IB models for different ages and \xeff\ values.  The right panel shows the
corresponding predictions for EB models.}
\label{mcorr}
\end{figure*}

\citet{MasHesse99b} showed that the star formation episodes taking place in 
{ compact} starburst or HII galaxies are generally of short duration, and that
their properties can be better reproduced by evolutionary synthesis models
assuming (nearly) instantaneous bursts, than by long-lasting, extended-in-time, star-formation processes. The generally strong optical emission lines in these
galaxies constrain the evolutionary state of their massive star-formation
episodes to ages below $6$ Myr, { typically within $4-5$ Myr.} On the
other hand, star formation is expected to proceed during long periods of time in
the disks of late-type spiral galaxies, generally in the form of individual
bursts at different times.  The accumulation of individual starburst episodes
along the disks of these galaxies mimick a continuous star formation process
\citep{MasHesse99b}, so that extended star formation models might be a better
approximation to reproduce their spatially-integrated properties. 

In Fig.~\ref{sxf-histo}, we have added to the predictions of the \lsxf\ values
the histogram corresponding to the whole sample, as well as the average
observational \lsxf\ value. As discussed above, there is a degeneracy between
\xeff\ and the age of the star formation episode, so that it is not possible to
discriminate between both parameters without additional constraints on the
evolutionary state.  { Apart from a few irregular galaxies, for which an instantaneous
model would describe better the star-formation episodes they are hosting, most
of the galaxies in the sample are large spiral galaxies experiencing long-term
star formation. Figure 5 shows that the mean \lsxf\ ratio can be reproduced by
relatively young EB models (after about 10 Myr of evolution) with moderate
efficiencies   \xeff\  of about $1$\%. Galaxies with higher  \lsxf\ ratios would
correspond to more evolved extended starbursts reaching the evolutionary
asymptotic  phase, with  \xeff\ in most cases below $10$\%.  On the other hand,
the galaxies with lower \lsxf\ values can only be reproduced by very young
models, after less than 10~Myr of evolution. The galaxies in the sample have been
selected in the various compilations for being {\em star-forming} or {\em
starburst} galaxies, i.e., galaxies experiencing a stronger than average episode
of star formation.  The integrated emission in the far infrared and soft X-ray
bands of some of these galaxies could be dominated  by a single but intense
burst of star formation. { While star formation proceeds in these objects for a
long time, these individual, intense bursts are not expected to last longer than
a few Myr.} Some of these starbursts could indeed be rather unevolved. We believe
that this is what we see in Fig.~\ref{sxf-histo}: some of the galaxies with low
\lsxf\ ratios could be dominated by a single, intense and relatively unevolved
burst of star formation. A deeper study of the individual galaxies would be
required to confirm this hypothesis, but it is beyond the scope of this work. 

\citet{Grimes05} computed the \lsxf\ ratios for a sample of
ultraluminous infrared, starburst and dwarf starburst galaxies (the galaxies
classified as starburst are already included in the  TUL06
sample).  For their 7 dwarf starburst galaxies (2 are already included in the
sample by RCS03), they derived a mean  \llsxf\ of $-4.0$, close
to the average value found for our complete compilation. The \lsxf\ ratio of
these galaxies (within the range $-3.58$ to $-4.29$) is properly reproduced by
IB models at approximately $4-5$ Myr, with \xeff\ within a realistic range  $1-5$\%
\citep{Strickland99,Summers01,Summers04}.  The mean \llsxf\ value of the 9
ultraluminous infrared galaxies (ULIRG) in their sample is $-4.5$, clearly below
our average.  This indicates that the star formation processes in this kind of
galaxy might be relatively unevolved and their emission dominated by a
single, intense episode of star formation. }

In Figure~\ref{mcorr}, we plot the observational \lsx\ vs. \lfir\ values
for the galaxies in the three samples. We include the predictions of
IB and EB models at different ages, and for \xeff\ values between $1$ and $10$\%.
{ These plots support the main points of our previous discussion: the observational \lsx\ vs.
\lfir\ correlation in star-forming galaxies can be well-reproduced by
evolutionary synthesis models, assuming realistic parameters: age below 6 Myr for
IB cases, and a spread of young to evolved bursts for EB models,  a high
efficiency in the reprocessing of UV photons into far infrared emission, and a
moderate ($1-10$\%)  efficiency in the heating of the diffuse interstellar
gas by the mechanical energy released by massive stellar winds. }

We have analyzed the effect of some intrinsic properties of the sampled objects,
on the dispersion shown in the correlation plots. First, as noted above, the
measured \lsx\ values did not include the correction for intrinsic neutral
hydrogen absorption. \citet{Kunth98} measured the column density of neutral
hydrogen in the line of sight to $8$ compact starburst galaxies, by fitting
their $Lyman\,  \alpha$ profiles, finding values in the range  $log(n_H) \sim
19 - 21$ cm$^{-2}$. We have computed that for the typical spectral properties of
the hot diffuse gas this correction would be within $0.4$\% -- $50$\% (i.e.
$<0.18$ dex). { A second effect would be the contamination of
the soft X-ray emission by low mass X-ray binaries associated to the underlying,
older stellar population. ROSA07 estimated that the contribution by LMXB
in their sample of spiral star-forming galaxies was generally of few percent,
and concluded that the contamination should be negligible for galaxies with
\sfr\ $ > 1 \, M_{\sun} \, yr^{-1}$. In the RCS03 sample only 7
objects have a value of \sfr\ below  $1 \, M_{\sun} \, yr^{-1}$, but their average \llsxf\
is $-3.7$ (in any case  within $-4.2$ to $-3.2$), i.e., with no deviation at all
from the total average \lsxf\ ratio. Therefore, the contamination of \lsx\ by
LMXB does not seem to be important. Finally, TUL06 removed the
contamination by point sources before computing the integrated \lsx, so that the
contamination of this sample by LMXB should be negligible. An additional effect
is related to the relative strength of the starburst emission compared to the
galaxy as a whole. While the soft X-ray emission would be associated mostly with
the starburst regions, the far infrared luminosity can include a significant
contribution from the rest of the galaxy, so that the \lsxf\ ratios obtained
from spatially-integrated measurements would be lower than the intrinsic value
produced by the starburst itself.  A detailed morphological analysis of the
galaxies in the sample is out of the scope of this paper, but we expect that at
least in some cases the observational \lsxf\ value might be contaminated by
emission not related to the star formation episode.}  

Finally, \lfir\ has been computed assuming almost complete reprocessing of UV
stellar continuum photons (\ebv\ $= 1$), as discussed in Sect.~2. {Lower
values of the extinction, of the order of \ebv\ $= 0.1$, would raise the predicted \lsxf\
ratios by up to $0.15$ dex.} The effective extinction of these objects should be
in between both extreme values {of \ebv.} In conclusion, correcting the soft X-ray
luminosity from intrinsic photoelectric absorption and/or rejecting the far
infrared emission not associated with the starburst regions themselves, could
increase the \lsxf\ ratios for some objects. On the other hand, smaller interstellar extinction values would increase the
predicted \lsxf\ ratio by less than $40$\%. 

We consider if it is possible to derive a calibration that would allow the \lsx\ luminosity to be used as a tracer of SFR. \citet{Strickland99} found from detailed hydrodynamical
simulations that superbubbles accelerated by the release of mechanical energy in
a starburst, would convert on average approximately $5$\% of the input mechanical energy
into soft X-ray emission. \citet{Summers01} analyzed Mrk~33, a dwarf
star-forming  galaxy, and concluded that it is dominated by an intense burst
$5$--$6$ Myr old, and that the rate of injection of mechanical energy from the
starburst is approximately $1.2\times10^{41}$ \ergs. The soft X-ray luminosity of the
central, extended diffuse gas derived by these authors is   $L_{\rm{soft
X}}\sim 2.2\times 10^{39}$ \ergs, corresponding to \xeff\ $\sim 0.018$. 
Similarly, \citet{Summers04} estimated the mechanical energy injection rate from
the starburst in NGC~5253 to be $L_{\rm{mech}} = 3.6\times10^{40}$ \ergs. The
measured thermal X-ray emission associated with the starburst was \lsx\ $\sim
4\times10^{38}$ \ergs, yielding \xeff\ $\sim 0.01$. 

{ We can therefore assume that  \xeff\ is constrained to be in the range
$1-5$\% for typical star-forming galaxies. Our evolutionary synthesis models
predict for  Instantaneous Bursts (of age between $3-6$ Myr) that \llsxf\ is approximately equal to \llsxf\ $\sim (-3.1, -4.0)$, with a central value
\llsxf\ $\sim -3.5$. On the other hand, Extended Burst models predict 
\llsxf\ $\sim (-3.0, -3.4)$ after $30$ Myr of evolution, when the number of
supernova explosions begins to stabilize.  For less-evolved, extended episodes
(of approximately 10 Myr of continuous star formation) the predicted ratios would be
lower, within the range \llsxf\ $\sim (-3.5, -4.0)$.

These values are close to the average ratio \llsxf\ $\sim -3.92$ derived from
our total  sample of star-forming galaxies, supporting the use of the soft X-ray
emission  as a tracer of the star formation rate in starburst galaxies. Using
the \lsx\ values predicted by the models, as shown in Fig.~\ref{lumin}, we can
derive the calibration of the star formation rate (or strength) as a function of
\lsx. For this we have considered the points at which  \llsxf\ $= -3.5$ for IB
models, \llsxf\ $= -3.1$ for EB cases in the asymptotic phase of evolution, and
\llsxf\ $= -3.7$ for extended, but non-evolved bursts, with ages of approximately 10~Myr, 
according to the discussion above. The calibrations would therefore be:  

\vspace{0.2cm}
\centerline  { \sfr\  (M$_{\sun}$  yr$^{-1})  =  2\times 10^{-41}$  \lsx\ (\ergs)  (evolved EB)}
 \vspace{0.2cm}

\centerline  { \sfr\  (M$_{\sun}$  yr$^{-1})  =  8\times 10^{-41}$  \lsx\ (\ergs)  (young EB) }
\vspace{0.2cm}

\noindent For comparison,  the semiempirical calibration derived by
\citet{Ranalli03}  is 

\vspace{0.2cm}
\centerline  { \sfr\  (M$_{\sun}$  yr$^{-1})  =  1.1\times 10^{-40}$  \lsx\ (\ergs) }
\vspace{0.2cm}

\noindent where we have  adapted the original calibration to the whole FIR
emission in the $1-1000 ~\mu$m band, to be consistent with  \citet{Kennicutt98}
(whose \lfir\ calibration is used as a proxy), and have scaled the original
\citet{Kennicutt98} mass normalization to the range $2-120 M_{\sun}$ to be
directly comparable with our results. 

\noindent These calibrations should be applicable, within the range of validity
shown in Fig.~\ref{lumin}, for galaxies experiencing an extended episode of
star formation at a constant star formation rate. In the case of coeval, instantaneous 
starbursts, the parameterization would be:    

\vspace{0.2cm}
\centerline  { \sfs\  (M$_{\sun}$)$  =  2\times 10^{-34}$  \lsx\ (\ergs) } 
\vspace{0.2cm}

\noindent where the star formation strength ($SFS$) is the total mass
transformed into stars at the onset of the burst. 

The semiempirical calibration proposed  by \citet{Ranalli03} is therefore in
good agreement with the calibration derived from synthesis models for relatively
unevolved star formation episodes, but it would tend to overestimate the star
formation rate for galaxies that have been forming massive stars over a long time, for example tens or hundreds of millions of years.

The models do not in principle predict any dependence of the \lsxf\ ratios
on the strength (i.e. total luminosity) of the star formation episodes, as
proposed by some authors for the hard X-ray luminosity. Nevertheless, we want
to remark that our models do not study the detailed properties of the
medium surrounding the star-forming regions. The intensity of the star-formation episode could, for example, have a direct effect on the dust-grain properties of the interstellar medium. \xeff\ may, in addition, be a function of the burst intensity. Both effects could create a correlation between \lsxf\ and the star-formation burst intensity, although such a correlation is unclear in Fig.~\ref{mcorr}. }

\section{Conclusions}

 We have compared the \lsxf\ values measured in a sample of  62 star-forming
galaxies with the predictions by our evolutionary synthesis models,  aiming to
analyze the validity of  semiempirical and theoretical calibrations of \lsx\ as a star
formation rate estimator. The main results can be summarized as follows:  

   \begin{enumerate}

\item The \lsxf\ ratios are strongly dependent on the efficiency in the
conversion of the mechanical energy released by the young massive starburst into
soft X-ray luminosity, by interaction of the stellar winds and supernova ejecta
with the surrounding interstellar medium. From theoretical predictions and
observational data, we expect an \xeff\ value of few percent in starburst
galaxies. 

\item \lsxf\ is also dependent on the evolutionary status of the star
formation episode. It increases rapidly with time during the first $5$ Myr of
evolution of a massive starburst, and shows a slower increase afterwards. After
a (nearly) instantaneous burst of star formation, \lsx\ decreases slower than
\lfir, as long as there remains a population of massive stars able to collapse
as supernovae (up to around 35~Myr).   

\item When star formation proceeds at a nearly constant rate during
extended periods of time, the \lsxf\ ratio is expected to stabilize after around
$40$ Myr, when the number of massive stars that produce supernovae has
reached an equilibrium between death and birth of new stars. 

{ 
\item The \lsxf\ values measured for the sample of star-forming galaxies are
consistent with the predictions by the models under realistic conditions:
relatively young and unevolved  star formation  episodes and \xeff\ values
within $1$--$10$\%.

\item A calibration of \lsx\ as a star formation rate estimator, based on the
predictions of evolutionary synthesis models, has been derived. The calibration
proposed by  \citet{Ranalli03} is consistent with the predictions for
relatively unevolved, time-extended bursts of star formation.   
}
\end{enumerate}

\begin{acknowledgements}

JMMH and HO are partially funded by Spanish MEC grants AYA2004-08260-C03-03 and
ESP2005-07714-C03-03. OH is funded by Spanish FPI grant BES-2006-13489. MCS
acknowledges funding by Spanish MEC grant AYA2004-02703, and by Spanish {\em
Ram\'on y Cajal} fellowship El 01/08/2007. 
\end{acknowledgements}

\end{document}